\documentstyle[prb,aps,preprint]{revtex}
\begin{document}
\draft
\title{The motion of holes on the triangular lattice: t-J model}
\author{Mohamed Azzouz}
\address{Centre de Recherche en Physique du Solide
et D\'epartement de Physique, Universit\'e de Sherbrooke
Qu\'ebec, J1K 2R1
Canada}
\author{Thierry Dombre}
\address{CNRS, Centre de Recherche sur les Tres Basses temp\'eratures,
BP. 166, 38042 Grenoble C\'edex9, France}
\date{\today}
\maketitle
\begin{abstract}
The motion of holes on the triangular lattice is studied using
the t-J model. Within the Born self-consistent approximation and the
exact Lanczos diagonalization, the single hole physics is first analyzed.
Then the spiral phase theory of Shraiman and Siggia is used to investigate
the case of a finite density of holes.
\end{abstract}
\section{introduction}
The main motivation for the recent innumerable works on the Hubbard
model is the Anderson suggestion \cite{anderson} that says
that this model, in its
simple version, can explain the physics of the CuO planes in high critical
temperature superconductors. The Hamiltonian of
this model is written as follows
\begin{equation}
H=-t\sum_{\langle i,j\rangle}\biggl(c_{i,\sigma}^{\dag} c_{j,\sigma}
+ h.c \biggr) + U \sum_i n_{i,\uparrow}n_{i,\downarrow}
\label{ham}
\end{equation}
where $n_{i,\sigma}=c_{i,\sigma}^{\dag}c_{i,\sigma}$ is the occupation
number of an electron with a spin $\sigma$ at the site $i$. The first
term of $H$ is the kinetic energy which allows for an electron to hop from
one site to one of its nearest neighbors with an amplitude $t$. The second
term stands for the on-site Coulomb repulsion ($U>0$) and
$c_{i,\sigma}^{\dag}$
and $c_{i,\sigma}$ are respectively creation and annihilation operators.

For large $U$, the double occupancy of a site with two electrons
$\uparrow$ and $\downarrow$ is energetically discarded. In
this limit the Hubbard model becomes equivalent to the so-called t-J model:
\begin{equation}
H_{t-J}=
P_0\biggl\{-t\sum_{\langle i,j\rangle}\biggl(c_{i,\sigma}^{\dag} c_{j,\sigma}
+ h.c \biggr)\biggr\}P_0 + J\sum_{\langle i,j\rangle}\biggl
({\bf S}_i\cdot{\bf S}_j -{1\over4}n_in_j\biggr)
\label{tJ}
\end{equation}
where
$P_0=\prod_i(1-n_{i,\uparrow}n_{i,\downarrow})$ applies the
constraint of non double occupancy.
${\bf S}={1\over2}\sum_{\alpha,\beta}
c_{i,\alpha}^{\dag}{\bf {\sigma}}_{\alpha,\beta}c_{i,\beta}$
is the spin operator ($\sigma^x$, $\sigma^y$ and $\sigma^z$ are the
Pauli matrices). The coupling constant of the magnetic part,
$J={{4t^2}/U}$, is positive.

At half filling (one electron per site on average), the first term of
$H_{t-J}$ is effectiveless and the model reduces to the Heisenberg model.
It is now widely believed that the ground state ($T=0K$) of
the Heisenberg model
on the square and triangular lattices presents antiferromagnetic
long rang order in the thermodynamic limit \cite{tangbernu}.

In this paper, the motion of a single hole is first studied on the
triangular lattice using the slave fermion representation
and an exact diagonalization. Then the collective instabilities
for a finite density of holes are investigated in the framework of a
mean-field approach first introduced by Shraiman and Siggia \cite{shraiman}
and known to lead to spiral phases. The main result in that part of our work is
that spiral phases in the doped case can only develop within the
preexisting plane of the $120^{\circ}$ AF structure of the pure
system. The discussion of the present results is made together
with a comparison with the square lattice results.

In the second section, the slave
fermion representation is used to write down an effective Hamiltonian from
(\ref{tJ}). Section III describes the self-consistent Born approximation
within which the single particle properties are calculated. The results,
analyzed in sections IV and V, distinguish between the cases
$J\ne0$ and $J=0$.
For the former case a {\it Fermi liquid} description can be done. However, for
the later one and for positive $t$, the quasi-particle vanishes
away from the centre of the Brillouin zone. The
question of a 'liquid' where
the quasi-particle depends on the position of the wavevector on
the Brillouin zone arises naturally and will be addressed elsewhere.
Section VI is devoted to
an exact calculation using the Lanczoc method. Good agreement concerning the
position of the energy minima as a function of ${\bf k}$ is obtained.
Finaly, the motion of a finite density of holes
is considered in sections VII and VIII where the spiral phase is calculated.

\section{The model}
\subsection{Slave fermion representation}
The t-J model is characterized by the existence of spin and charge
degrees of freedom. In one-dimension (1D), these degrees of freedom
separate and one ends with a free propagation of the hole and the domain wall
in the AF background. This illustrates the mechanism
of spin-charge separation in 1D. In 2D, a string of unsatisfied bonds
takes place behind the moving hole. Spin and charge degrees {\it remain
coupled}.

The constrained electronic operator is written as follows
\begin{equation}
c_{i,\sigma}(1-n_{i,\sigma})\equiv\tilde{c}_{i,\sigma}
=\psi_i^{\dag}b_{i,\sigma}
\label{c}
\end{equation}
where $\psi_i^{\dag}$ is a fermionic operator which creates a spinless
hole at the site $i$ and $b_{i,\sigma}^{\dag}$ is the Schwinger
boson \cite{schwinger} operator which creates
a spin $\sigma$ at $i$. The number of bosons and fermions
on each site must satisfy the following constraint
\begin{equation}
\psi_i^{\dag}\psi_i+\sum_{\sigma}b_{i,\sigma}^{\dag}b_{i,\sigma}=2S
\label{constraint}
\end{equation}
The spin operator is written in the form
\begin{equation}
{\bf \vec S}_i=\chi_i^{\dag}{\bf \vec \sigma}\chi_i,
\label{spin}
\end{equation}
where
\begin{equation}
\chi_i=\pmatrix{b_{i,\uparrow}\cr
b_{i,\downarrow}}
\label{spinor}
\end{equation}
is a two-component spinor. The Schwinger bosons formalism
is easily generalized
to higher values of the spin $S$ where an expansion on $1/2S$ allows one
to recover the spin wave theory, that is
\begin{equation}
\chi_i=\pmatrix{\sqrt{2S-b_i^{\dag}b_i}\cr
b_i}
\approx\pmatrix{\sqrt{2S}-b_i^{\dag}b_i/2\sqrt{2S}\cr
b_i}
\label{swt}
\end{equation}
for large $S$. Here $b_i$ is the Holstein-Primakoff bosonic operator
describing small deviations around the classical spin configuration which
minimizes the energy (this configuration is hereafter called classical
N\'eel state as on the square lattice):
\begin{equation}
\chi_i=\pmatrix{2S\cr
0}
\label{neel}
\end{equation}
The constraint (\ref{constraint}) is transformed into a non holonomic
constraint
\begin{equation}
\psi_i^{\dag}\psi_i+b_i^{\dag}b_i\le 2S
\label{constraint2}
\end{equation}
which is more difficult to handle than (\ref{constraint}). The latter
is taken into account by introducing a site dependent Lagrange multiplier
$\lambda_i$ and the effective action is written as follows
\begin{eqnarray}
{\cal S}=\int_0^\beta{\rm d}\tau\biggl\{
&&-t\sum_i\psi_i^{\dag}\partial_{\tau}\psi_i
+ \sum_i b_{i,\sigma}^{\dag}\partial_{\tau}
b_{i,\sigma} -t\sum_{\langle i,j\rangle}
\psi_i^{\dag}\psi_jb_{j,\sigma}b_{i,\sigma}^{\dag}\cr
&&+J\sum_{\langle i,j\rangle}\biggl({1\over2}\chi_i^{\dag}{\bf \sigma}
\chi_i\biggr)\cdot\biggl({1\over2}\chi_j^{\dag}{\bf \sigma}
\chi_j\biggr)\cr
&&+\sum_i\lambda_i\biggl(\psi_i^{\dag}\psi_i+\chi_i^{\dag}\chi_i
-2S\biggr)\biggr\}.
\label{hamiltonian}
\end{eqnarray}
In the spin wave theory the constraint is neglected and we do so
in the present work.

\subsection{Effective Hamiltonian on the triangular lattice}
The classical N\'eel state on the triangular lattice is displayed in
Fig.\ \ref{fig1}. The spins are oriented, on a coplanar configuration,
in such a way that two adjacent spins make
an angle of $2\pi/3$. These different orientations define three sublattices
A, B and C instead of two sublattices as it is the case on the square
lattice where two adjacent spins are antiparallel. By carrying
out a local rotation of $2\pi/3$ and $-2\pi/3$ on the sublattices B and C
respectively, that is
\begin{equation}
\chi_{B(C)}={\cal R}_{B(C)}\pmatrix{\sqrt{1-b^{\dag}b}\cr ib}
\label{rotation}
\end{equation}
where ${\cal R}_{B(C)}$ is the SU(2) matrix representing the rotation on
B (C), the system is made formally ferromagnetic but the resulting
Hamiltonian keeps naturally its AF character:
\begin{eqnarray}
H_{t-J}\to H=&&-{t\over2}\sum_{\langle i,j\rangle}(\psi_i^{\dag}\psi_j+hc)
-{t\over2}\sum_{\langle i,j\rangle}(\psi_j^{\dag}\psi_jb_j^{\dag}b_i+hc)\cr
&&-{{t\sqrt{3}}\over2}\biggl(\sum'\psi_i^{\dag}\psi_j(b_i-b_j^{\dag})-
\sum''\psi_i^{\dag}\psi_j(b_i-b_j^{\dag})\biggr)\cr
&&+{{3J}\over2}\sum_ib_i^{\dag}b_i+{J\over8}\sum_{\langle i,j\rangle}
\biggl(b_i^{\dag}b_j+b_j^{\dag}b_i-3(b_ib_j+b_j^{\dag}b_i^{\dag})\biggr).
\label{htr}
\end{eqnarray}
The hole moves as a free particle (coherent motion) through the first term in
(\ref{htr}) which is present since the spinors on adjacent sites are not
orthogonal. In the case of the square lattice, such a term is absent.
The second term in (\ref{htr}) allows an exchange of a spin flip and the hole
by conserving the total number of overturned spins. In the third term, there
is a change in the sign depending on wether the spin flip is absorbed or
created after the hopping of the hole. $\sum'$ ($\sum''$) refers to
the propagation of the hole in the
three directions ${\bf e}_i$ ($-{\bf e}_i$), Fig.\ \ref{fig1}.
The last term is the AF exchange interaction written in spin wave theory.

The presence of coherent and incoherent propagations of the hole is a
salient feature of the triangular antiferromagnetic lattice, which in
some sense interpolates between the ferromagnetic and antiferromagnetic
limits on the square lattice. In the last case, only incoherent
propagation is possible in the absence of quantum spin fluctuations. Arguments
taken from the Brinkman-Rice picture\cite{brinkman}
shows that the most important hopping terms in (\ref{htr})
are the first and the third. The point is that the
coherent propagation alone would give a minimum in the hole dispersion
of $-3t$ for $t>0$ or of $3t\over 2$ for $t<0$. But the hole can still
lower its kinetic energy in a quite significant manner by using the
incoherent channel. Indeed, would the third term be alone, then the estimate
$-t \sqrt {15}$ obtained in the self-retracing path approximation would
apply. On the other hand, we expect the second term to play only a
minor role and shall therefore neglect it in the following.

Transcribed in ${\bf k}$ space, Eq.\ (\ref{htr}) yields
\begin{eqnarray}
H=&&-it{\sqrt{3}\over N}\sum_{{\bf k},{\bf q}}\psi_{{\bf k}+{\bf q}}^{\dag}
\psi_{\bf k}
\biggl[(u_{\bf q}h_{\bf k}-v_{\bf q}h_{{\bf k}+{\bf q}})\alpha_{\bf q}+
(v_{\bf q}h_{\bf k}-u_{\bf q}h_{{\bf k}+{\bf q}})\alpha_{-{\bf q}}^{\dag}
\biggr]\cr
&&-t\sum_{\bf k}\gamma_{\bf k}\psi_{\bf k}^{\dag}\psi_{\bf k}
+J\sum_{\bf q}\omega_{\bf q}\alpha_{\bf q}^{\dag}\alpha_{\bf q}
\label{hk}
\end{eqnarray}
where
$$
\gamma_{\bf k}=\sum_{{\bf e}_i}\cos({\bf k}\cdot{\bf e}_i),\
h_{\bf k}=\sum_{{\bf e}_i}\sin({\bf k}\cdot{\bf e}_i)
$$
and
$$
\omega_{\bf q}={3\over2}J\sqrt{(1-{\gamma_{\bf q}\over3})
(1+{{2\gamma_{\bf q}}\over3})}.
$$
The operator $\alpha_{\bf q}$ obtained from $b_{\bf q}$ by using the
Bogolubov transformation, creates a spin wave with a wavevector ${\bf q}$.
$u_{\bf q}$ and $v_{\bf q}$ are the coherence factors:
$$
u_{\bf q}=\sqrt{{J\over{2\omega_{\bf q}}}}\sqrt{{3\over2}
+{\gamma_{\bf q}\over4}+{{\omega_{\bf q}}\over J}}
$$
$$
v_{\bf q}={\rm sign}(-\gamma_{\bf q})
\sqrt{{J\over{2\omega_{\bf q}}}}\sqrt{{3\over2}
+{\gamma_{\bf q}\over4}-{{\omega_{\bf q}}\over J}}
$$
In the following, (\ref{hk}) is investigated using the self consistent
Born approximation as developped in \cite{horsch}.

\section{self-consistent Born approximation}

The single particle Green's function is defined as follows
\begin{equation}
G_{\sigma}({\bf k},\omega)=\langle \Phi_0|c_{{\bf k},\sigma}^{\dag}
{1\over{\omega-H}}c_{{\bf k},\sigma}|\Phi_0\rangle
\label{green1}
\end{equation}
where $|\Phi_0\rangle$ is the ground state of the AF Heisenberg part of the
Hamiltonian. Using (\ref{c}) one gets
\begin{equation}
c_{k,\sigma}={1\over{\sqrt{N}}}\sum_{k'}\psi_{k'}^{\dag}b_{k+k',\sigma}
\label{equation}
\end{equation}
for the electronic operator in the slave fermion representation. As an
approximation, the operator $b_{k+k',\sigma}$ is replaced by its
mean value in the
N\'eel configuration since we consider that the important physics is contained
in the fermionic part $\psi$. We define the Green's function related to
$\psi$ by
\begin{equation}
G({\bf k},\omega)=\langle \Phi_0|\psi_k
{1\over{\omega-H}}\psi_k^{\dag}|\Phi_0\rangle.
\label{green2}
\end{equation}
In the N\'eel state, we have
$$
\langle b_{i,\uparrow}\rangle\approx \exp{(2im\pi/3)},
\ {\rm and}\
\langle b_{i,\downarrow}\rangle\approx \exp{(-2im\pi/3)}
$$
where $m$ takes $0$, $1$ and $-1$ on the three sublattices A, B and C
respectively. This can be written in the form
$$
\langle b_{i,\uparrow}\rangle\approx \exp{(-i{\bf G}\cdot{\bf R}_i)},
\ {\rm and}\
\langle b_{i,\downarrow}\rangle\approx \exp{(i{\bf G}\cdot{\bf R}_i)}
$$
where
${\bf G}=(4\pi/3){\bf x}$. Transcribed in Fourier space, these yield
$$
\langle b_{k,\uparrow}\rangle\approx \sqrt{N}\delta_{{\bf k},-{\bf G}},
\ {\rm and}\
\langle b_{k,\downarrow}\rangle\approx \sqrt{N}\delta_{{\bf k},{\bf G}}.
$$
A simple correspondance between the
true Green's function, (\ref{green1}), and the $\psi$-Green's function,
(\ref{green2}), namely :
\begin{eqnarray}
&&G_{\uparrow}({\bf k},\omega)\approx G(-{\bf k}-{\bf G},\omega)\cr
&&G_{\downarrow}({\bf k},\omega)\approx G(-{\bf k}+{\bf G},\omega)
\label{corres}
\end{eqnarray}
allows the evaluation of the true electronic Green's function $G_\sigma$
using $G$. In this section, we concentrate on the Green's function of
the slave fermion.

The Dyson equation for $G({\bf k},\omega)$ is
\begin{equation}
G({\bf k},\omega)={1\over{\omega-\epsilon({\bf k})-\Sigma({\bf k},\omega)}}
\label{dyson1}
\end{equation}
where $\epsilon({\bf k})=-t\gamma({\bf k})$ is the free part of kinetic
energy and $\Sigma({\bf k},\omega)$ is the self energy resulting from
the incoherent motion of the hole. The evaluation of $\Sigma({\bf k},\omega)$
is done using the diagram of Fig.\ \ref{fig2} which corresponds to the
following expression
\begin{equation}
\Sigma({\bf k},\omega)={{3t^2}\over N}\sum_{\bf q}|M({\bf k},{\bf q})|^2
G^{(0)}({\bf k}+{\bf q},\omega-\omega({\bf q}))
\label{self1}
\end{equation}
where $G^{(0)}$ is the $\psi$-Green's function of the free part of
the Hamiltonian.
The self-consistent approximation consists in replacing $G^{(0)}$ in
(\ref{self1}) by the total Green's function $G$. In this approximation,
(\ref{dyson1}) becomes
\begin{equation}
G({\bf k},\omega)=\biggl(\omega-\epsilon({\bf k})-
{{3t^2}\over N}\sum_{\bf q}|M({\bf k},{\bf q})|^2
G({\bf k}+{\bf q},\omega-\omega({\bf q}))\biggr)^{-1}
\label{dyson2}
\end{equation}
where $M({\bf k},{\bf q})=v_qh_k-u_qh_{k+q}$.

In this approach, a series of an infinite number of diagrams which
neglect vertex corrections is considered as shown in Fig.\ \ref{fig2}.
Vertex corrections roughly correspond to Trugman processes \cite{trugman},
which are known to become important fot $t\gg J$.
In the most elementary process of this type on the triangular lattice, the
hole makes one turn and a half around a triangular plaquette, which
gives rise to coherent nearest-neighbor hopping. Trugman processes tend
to push the energy minima towards the corners of the hexagonal
Brillouin zone (independent of the sign of $t$). Their influence should
remain week for moderate values of $J\over t$, as on the square lattice.

\section{results}

Eq.\ (\ref{dyson2}) is computed by means of an iterative method. We
fixe an arbitrarily initial function $G({\bf k},\omega)$ for all wavevectors
${\bf k}$, belonging to the hexagonal Brillouin zone, and frequencies $\omega$,
and use (\ref{dyson2}) to iterate until convergence towards
the unique solution. The unicity
has been proved by starting from different initial functions.

The numerical calculation is performed for hexagonal clusters where the
sites are labelled as shown in Fig.\ \ref{fig3} and parametrized as follows :
\begin{eqnarray}
&&k_x={{2\pi}\over{3\ell}}m,\     0\le m<3\ell\cr
&&k_y={{4\pi}\over{\sqrt{3}\ell}}\biggl(n+{1\over2}{{1-(-1)^m}\over2}\biggr),\
0\le n<\ell
\label{parametrization}
\end{eqnarray}
to get the right number of independent sites inside the first Brillouin zone.
The results presented here are from a cluster of $108$ sites ($\ell=6$).
The number of independent sites is $3\ell^2$.

The quantity of interest is the spectral function which is related to
the Green's function through the relation
\begin{equation}
{\cal A}({\bf k},\omega)=-{1\over\pi}{\rm Im} G({\bf k},\omega).
\end{equation}
At finite $J$, ${\cal A}({\bf k},\omega)$
presents a ${\bf k}$-dependent
quasi-particle peak at low energies. This is consistent with a Fermi liquid
picture for positive and negative $t$. Note that because the electron-hole
symmetry is absent for the t-J model on the triangular
lattice, the spectral function
depends on the sign of $t$. This is not the case on the square lattice
where there is a $t\to -t$ symmetry.

For negative $t$ and finite $J$, the energy minimum of the
quasi-particle peak is
located at ${\bf k}=(\pi,\pi/\sqrt{3})$ which corresponds to the middle of
the edges of the Brillouin zone. For positive $t$, however the minimum
is realized at the centre of the Brillouin zone: ${\bf k}=(0,0)$. The spectral
functions of these wavevectors, together with those of ${\bf k}=(4\pi/3,0)$,
are displayed in Fig.\ \ref{fig4} and Fig.\ \ref{fig5} for $J/|t|=0.2$.
\section{properties of the quasi-particle}
\subsection{Spectral weight of the quasi-particle for $t<0$}
In this section, we are interested in analyzing the properties of
the quasi-particle in the case of negative $t$.
The spectral weight $a({\bf k})$ of the quasiparticle is calculated by
computing the area under the quasiparticle peak. The results for
small values of $J$ suggest a simple power law as a function of $J$:
\begin{equation}
a({\bf k})\simeq\beta J^\alpha
\label{weight}
\end{equation}
where the values of $\alpha$ and $\beta$ are summarized in table I for
the three characteristic wavevectors ${\bf k}=(0,0)$, $(4\pi/3,0)$ and
$(\pi,\pi/\sqrt{3})$.

Contrary to the case of free electrons where
the weight is a constant ($=1$), $a({\bf k})$ is here ${\bf k}$-dependent.
The zero quasi-particle weight at $J=0$ implies the absence
of the quasiparticle peak at low energies as shown by the spectral
function. What the absence of quasi-particle means is the
breakdown of the Fermi liquid picture. However, Nagaoka theorem \cite{hal}
applies for negative $t$ and ensures that the ground state is ferromagnetic
at $U=\infty$ ($J=4t^2/U$). Our result in this case is an artifact
of the approximation
which states an AF background. For finite $J$, it is natural to consider
such a background and our results are physically meaningful.

For $J\gg|t|$, the spectral weight goes to unity since
$$
a({\bf k})={1\over{1-{\partial \over{\partial \omega}}
{\rm Re}\Sigma({\bf k},\omega)}}\ ,\ \omega=E({\bf k})
$$
and
$$
{\partial \over{\partial \omega}}
{\rm Re}\Sigma({\bf k},\omega)\simeq-{{3t^2}\over N}\sum_{\bf q}
{{|M({\bf k},{\bf q})|^2}\over{(\omega({\bf q})-t\gamma_{\bf k})^2}}
\sim {{t^2}\over{J^2}}
$$
\subsection{quasi-particle dispersion relation for $t<0$}
By performing the numerical calculation of the position of the quasi-particle
peak for every wavevector belonging to the Brillouin zone, we get the
dispersion in ${\bf k}$ of the energy minimum. This energy is well fitted
by the following simple expression
\begin{equation}
E({\bf k})=A+B\gamma_{\bf k}+C{h_{\bf k}}^2
\label{dispersion}
\end{equation}
suggested by the arguments developped in \cite{azzouz}. $A$, $B$ and $C$ are
$J$-dependent as seen in table II.

This fit contains hopping processes to nearest neighbors through the
$B$-term and second nearest neighbor through the $C$-term.
$B\gamma_{\bf k}$ comes from the coherent part of the Hamiltonian, whereas
$C{h_{\bf k}}^2$ originates from the incoherent part with the partcipation
of quantum spin fluctuation. So at finite small $J$, a Fermi liquid picture is
appropriate to describe the motion of the hole as a quasi-particle with
the dispersion relation given by (\ref{dispersion}).

For $J\gg |t|$, one gets an analytical expression for $E({\bf k})$, namely
\begin{equation}
E({\bf k})\simeq -t\gamma_{\bf k}-\sum_{\bf q}
{{|M({\bf k},{\bf q})|^2}\over{\omega({\bf q})-t\gamma_{{\bf k}+{\bf q}}}}.
\label{dispersion1}
\end{equation}

\subsection{The case of $t>0$}

At finite $J$ and positive $t$, the physics is similar to that of
the case of $t<0$. But for $J=0$, the situation is quite different since
the Nagaoka theorem is not satisfied and the
energy of the hole can be minimized further in a singlet spin state. It is
then natural to consider the AF background as a good approximation
for $t>0$ even when J=0. From the calculation of the spectral
function ${\cal A}({\bf k})$,
we found out that the energy minimum is located at ${\bf k}=(0,0)$. For
${\bf k}=(0,0)$ ($J=0$), a well defined quasi-particle peak is present.
However, the quasi-particle is strongly ${\bf k}$-dependent.
the peak loses in intensity and broadnes, as illustrated in Fig.\ \ref{fig5},
when we run away from the center of the Brillouin zone.
Therefore, depending on the wavevector, a Fermi
liquid interpretation, {\it in the classical way}, is either possible or not.
We believe that Trugman
processes will not spoil this conclusion since they only renormalize
(slightly) the coherent part of the hole motion.
This discussion rises the question of an electronic system where the
quasi-particle weight depends on the wavevector ${\bf k}$ and vanishes
on some points of the Brillouin zone \cite{azzouz1}. A work concerning this
question is in progress.

\section{Exact diagonalization}
The exact diagonalization is performed using the Lanczos method on a
$3\times2^2$ hexagonal cluster (Fig.\ \ref{fig3}). The dispersion relation and
the total
spin are calculated.

The results for $t<0$ and $t>0$ are presented. For $t<0$ the dispersion
relation $E({\bf k})$ and the total spin $S_{tot}$ are summarized in
tables III and IV respectively. $E({\bf k})$ is reported
as a function of the wavevector ${\bf k}$ and $J/|t|$. First let us
consider the case $t<0$. For $0\le J \le0.3$, the energies
have a small dependence on ${\bf k}$. But for $J>0.3$, the energy minimum
is located at ${\bf k}=(2\pi/3,0)$ which transforms into the middle of
the Brillouin zone sides, namely ${\bf k}\equiv(\pi,\pi/\sqrt{3})$, by
a translation ${\bf G}=(4\pi/3,0)$. This is consistent with what is obtained
in the approach of the previous sections. However, a translation of $-{\bf G}$
produces a meaningless result. The reason for this discrepancy is due
to the fact that in the slave fermion approach the chiral symmetry is broken,
whereas in the exact diagonalization the symmetry cannot be
spontaneously broken.

Things become more understandable if we calculate the total spin $S_{tot}$.
What is clear from the table IV is that for $J>0.3$,
$S_{tot}$ is small for
all the values of ${\bf k}$ and the ground state is a singlet: $S_{tot}=1/2$.
For $0\le J\le0.3$, $S_{tot}$ is big: $5/2\le S_{tot}\le11/2$. For $J=0$,
the ground state, given at ${\bf k}=(0,0)$, is ordered
ferromagnetically, a result which agrees with the
Nagaoka theorem which applies only to negative $t$ on the triangular
lattice.
For $0\le J\le0.3$ and ${\bf k}\neq(0,0)$, the energies are very close to $-6$
(the lowest energy). The ferromagnetic states have, however
the energies $2$, $-1$ and $3$ (in the units of $|t|$) for
$(\pi,\pi/\sqrt{3})$, $(2\pi/3,0)$ and $(4\pi/3,0)$ respectively.
A possible explanation is that the system finds a compromise
in which the spins deviate slightly from the ferromagnetic state
(the spins do not remain in coplanar positions) to keep
a high value of $S_{tot}$ but the energies are close to $-6$.
This phenomenon is similar to Aharonov-Bohm effect \cite{aharanov}.
The stability
of this phase as the size of the cluster increases has to be clarified.
On the other hand, it is natural to think that the AF correlations, growing
with the cluster size, will rise above any other correlations at $J>0$.

For large $J$, the antiferromagnetic correlations become dominant and the
total spin becomes small. In this case the exact results of the minimum
of the dispersion relation compare well with the self-consistent approach.

The situation is simpler for positive $t$. The results are summarized in tables
V and VI. There is no crossing in the energy
levels as it is the case for $t<0$ (table III). The minimum is always
located at
${\bf k}=(4\pi/3,0)$ which transforms into ${\bf k}=(0,0)$ using
Eq.\ (\ref{corres}).
The total spin of the ground state is $S_{tot}=1/2$. Here also a good
agreement, concerning the minimums of $E({\bf k})$,
between the two approaches is obtained.
For $J=0$, the fact that the ground state is not ferromagnetic confirms
that the Nagaoka theorem does not apply for $t>0$.

\section{motion of a finite density of holes}

We assume that the holes live around the minima calculated in
the case of one hole. They form small pockets or valleys, whose area
is proportional to their density. We shall use the slave fermion picture from
now on. There is only one valley around the centre
of the Brillouin zone for $t>0$ but three of them on the edges of the
Brillouin zone for $t<0$. We are interested in the long-range
interaction between holes, as mediated by low energy spin-waves.

The expression of the coupling of holes to the spin waves is
\begin{equation}
-it{\sqrt{3}\over N}\sum_{{\bf k},{\bf q}}\psi_{{\bf k}+{\bf q}}^{\dag}
\psi_{\bf k}
\biggl[(u_{\bf q}h_{\bf k}-v_{\bf q}h_{{\bf k}+{\bf q}})\alpha_{\bf q}+
(v_{\bf q}h_{\bf k}-u_{\bf q}h_{{\bf k}+{\bf q}})\alpha_{-{\bf q}}^{\dag}
\biggr].
\label{swhole}
\end{equation}
Among the three Goldstone modes (${\bf q}=0$ and ${\bf q}=\pm(4\pi/3,0)$),
only ${\bf q}=0$ is relevant since it is easily seen that two valleys cannot
be coupled by a momentum transfer ${\bf q}=\pm(4\pi/3,0)$, Fig.\ \ref{fig6}
in the case
$t<0$. This is {\em a fortiori} true in the case $t>0$ where there is only
one valley left. For ${\bf q}\sim0$,
(\ref{swhole}) becomes
\begin{eqnarray}
-it{\sqrt{3}\over {2N}}&&\sum_{{\bf k},{\bf q}}\psi_{{\bf k}+{\bf q}}^{\dag}
\psi_{\bf k}\biggl[h_k(u_q-v_q)(\alpha_{q}-\alpha_{-q}^{\dag})\biggr]\cr
&&-it{\sqrt{3}\over {4N}}\sum_{{\bf k},{\bf q}}\psi_{{\bf k}+{\bf q}}^{\dag}
\psi_{\bf k}\biggl[(u_q+v_q)(-{\bf q}.{\bf {\nabla}}h_k)
(\alpha_q+\alpha_{-q}^{\dag})\biggr]
\end{eqnarray}
which transforms into
\begin{equation}
-it{\sqrt{3}\over {2N}}\sum_{{\bf k},{\bf q}}\psi_{{\bf k}+{\bf q}}^{\dag}
\psi_{\bf k}h_k(-iS^z(q))
+t{\sqrt{3}\over {2N}}\sum_{{\bf k},{\bf q}}\psi_{{\bf k}+{\bf q}}^{\dag}
\psi_{\bf k}({\bf p}_k.(i{\bf q}\phi({\bf q})))
\label{phe}
\end{equation}
where
\begin{equation}
S^z({\bf q})=i(b_q-b_{-q}^{\dag})/2
\end{equation}
is the slowly varying component of the magnetization in the $z$-direction,
while
\begin{equation}
\ \phi({\bf q})=(b_q+b_{-q}^{\dag})/2
\end{equation}
parametrizes the slow distortion of the ordered $120^{\circ }$ sructure
within its plane. The vectorial quantity
\begin{equation}
{\bf p}_k={\bf {\nabla}}_kh_k
\end{equation}
appears like the effective dipolar momentum carried by the hole.
It can be shown that the coupling to $S^z({\bf q})$, which semiclassically
behaves as $\partial \phi \over \partial t$, does not lead to long-range
interaction between holes. On the other hand, the second term, once
written in real space, becomes
\begin{equation}
t{\sqrt{3}\over {2N}}\sum_{\bf r}\sum_i
\psi_i({\bf r})^{\dag}\psi_i({\bf r})({\bf p}_i.{\bf \nabla}_{\bf r}
\phi({\bf r}))
\end{equation}
where the $i$ indicates the different valleys (see Fig.\ \ref{fig6}).
The effective dipolar momentum can safely be taken equal to its value at the
centre of each valley for small hole density. Two cases are to be considered:\\
{\it i}) $t>0$: the minimum of the dispersion relation is located at
${\bf k}=0$. There is only one valley and the dipolar momentum ${\bf p}=0$.
So, no dipolar interaction between the holes and the spin waves can be
generated. The interaction is at least quadrupolar and decays as $r^{-4}$.
{\it ii}) $t<0$: the minima of the dispersion relation are located at
${\bf k}=(\pi,\pi/\sqrt{3})$ and equivalent momenta, Fig.\ \ref{fig6}.
The dipolar momenta in this case do not vanish. They take the following values
\begin{equation}
{\bf p}_1=2{\bf x},\ {\bf p}_2=-{\bf x}+\sqrt{3}{\bf y},\
{\bf p}_3=-{\bf x}-\sqrt{3}{\bf y}.
\end{equation}

\section{spiral phases}
According to the results of the previous sections, we introduce
a phenomenological Hamiltonian for the motion of a finite
density of holes. The spatial density of this Hamiltonian is given by:
\begin{equation}
{\cal H}=\sum_{i=1,2,3}\biggl\{-{1\over2}\psi_i^{\dag}{\bf \nabla}^2\psi_i
+ga\psi_i^{\dag}\psi_i ({\bf p}{\bf \nabla})\phi\biggr\}
+{1\over2}\bigl(\rho({\bf \nabla}\phi)^2 + {M^2\over\chi}\bigr)
\label{holes}
\end{equation}
where $M$ is the z-component of the magnetization ($M\sim0$). The
coupling constant
$g$ is of the order of $t$ for $t\ll J$ ($g=ta\sqrt{3}/4$). When $J\ll t$,
the vertex corrections become important and are expected to renormalize
$g$ to an order of $J$. Using the expressions of ${\bf p}_1,{\bf p}_2,
{\bf p}_3$ given above, it can be seen that
$\partial_x\phi$ and $\partial_y\phi$ are respectively coupled to
$2n_1-n_2-n_3$ and $\sqrt{3}(n_2-n_3)$ where
$n_i=\langle\psi_i^{\dag}\psi_i\rangle$ is the hole density in the valley
$i$. At the mean-field level, we can write:
\begin{equation}
{\cal H}_{int}=ga{(\rm Re}\biggl\{n_1+n_2{\rm e}^{-2i\pi/3}
+n_3{\rm e}^{2i\pi/3})(\partial_x\phi+i\partial_y\phi)\biggr\}
+|\partial_x\phi+i\partial_y\phi|^2.
\label{interaction}
\end{equation}
The minimization with respect to $\phi$ yields
\begin{equation}
\partial_x\phi+i\partial_y\phi=-{{ga}\over\rho}(n_1+n_2{\rm e}^{-2i\pi/3}
+n_3{\rm e}^{2i\pi/3}).
\label{moyen}
\end{equation}
Eq.\ (\ref{interaction}) becomes
\begin{equation}
{\cal H}_{int}=-{1\over2}{{g^2a^2}\over\rho}
|n_1+n_2{\rm e}^{-2i\pi/3}+n_3{\rm e}^{2i\pi/3}|^2.
\end{equation}
and the kinetic energy can be written as follows
\begin{equation}
{\pi\over m}(n_1^2+n_2^2+n_3^2)={\pi\over{3m}}\biggl((n_1+n_2+n_3)^2
+2|n_1+n_2{\rm e}^{-2i\pi/3}+n_3{\rm e}^{2i\pi/3}|^2\biggr)
\end{equation}
so that the normal phase becomes unstable when
$$
{{2\pi}\over{3m}}-{1\over2}{{g^2a^2}\over\rho}<0
$$
that is when ${3g^2a^2m/\rho\pi}$. Once this condition is satisfied,
the system maximizes
$$
|n_1+n_2{\rm e}^{-2i\pi/3}+n_3{\rm e}^{2i\pi/3}|^2=
{1\over2}\biggl((n_1-n_2)^2+(n_2-n_3)^2+(n_3-n_1)^2\biggr).
$$
This is realized for $n_1=n>0$ and $n_2=n_3=0$ or all the solution obtained
by cyclic permutation of the indices 1, 2 and 3. This implies that only
one valley is occupied and the two others are empty in the spiral phase.
A simple interpretation of this phase in the real space is obtained using
Eq.\ (\ref{moyen}). For the phase $n_1=n$ and $n_2=n_3=0$ one has
$$
\langle\partial_x\phi\rangle=-{{ga}\over\rho}n,\
{\rm and}\ \langle\partial_y\phi\rangle=0,
$$
which means that the spins rotate around their position in the normal phase
uniformly when moving along the $x$-axis. In the other solutions for $n_i$,
the rotation occurs along the axes ${\bf e}_2$ and ${\bf e}_3$
for $n_1=n_3=0$ and $n_2=n$, and $n_1=n_2=0$ and $n_3=n$ respectively.


%
\acknowledgements
We are grateful to J. C. Angles D'Auriac for helpful and
interesting discussions. One of us (M. A) would like to thank
the Natural Sciences and
Engineering Research Council of Canada (NSERC) and the Fonds pour
la formation de chercheurs et l'aide a la recherche from the Government of
Qu\'ebec (FCAR) for financial supporting.
\begin{figure}
\caption{The {\it classical N\'eel state} in the case of the
triangular lattice is drawn on one plaquette. The three sublattices A, B and C
are shown as well. ${\bf e}_i$ correspond respectively to the vectors on A, B
and C sublattices for $i=1$, $i=2$ and $i=3$.}
\label{fig1}
\end{figure}

\begin{figure}
\caption{On the left hand side of the figure, the diagram used for
self-energy is shown (with the bare propagator substituted by the true one).
The expansion in the bare propagator is shown on the right hand side. The
wavy line refers to the spin-wave propagator.}
\label{fig2}
\end{figure}

\begin{figure}
\caption{The Brillouin zone is displayed for $\ell=2$. In this
case, the wavevectors ${\bf k}=(0,0)$, $(\pi,\pi/3^{1/2})$
and $(4\pi/3,0)$
correspond respectively to the points $(m,n)=(0,0)$, $(3,0)$ and $(4,0)$.}
\label{fig3}
\end{figure}

\begin{figure}
\caption{The spectral functions of: (a) $(0,0)$, (b) $(4\pi/3,0)$ and (c)
$(\pi,\pi/3^{1/2})$ are presented as a function of frequency. $\ell=6$,
$t=-1$ and $J=0.2|t|$.}
\label{fig4}
\end{figure}

\begin{figure}
\caption{The spectral functions of: (a) $(0,0)$, (b) $(4\pi/9,0)$ and (c)
$(\pi,\pi/3^{1/2})$ are presented as a function of frequency. $\ell=6$,
$t=-1$ and $J=0$. ${\bf k}=(4\pi/9,0)$ is located near the point $(2,0)$
of Fig. 3. The quasi-particle broadens.}
\label{fig5}
\end{figure}

\begin{figure}
\caption{The position of the valleys and their dipolar momenta $p_i$
are presented on the Brillouin zone for $i=1$, $i=2$ and $i=3$.}
\label{fig6}
\end{figure}
\newpage
\begin{table}
\caption{The values of the coefficients $\alpha$, and $\beta$ of
Eq. (23).}
\end{table}

\begin{table}
\caption{The values of the coefficients $A$, $B$ and $C$ of
Eq. (24).}
\end{table}

\begin{table}
\caption{The hole energy as a function of ${\bf k}$ and $J$ for negative
$t(=-1)$}
\end{table}

\begin{table}
\caption{Total spin as a function of ${\bf k}$ and $J$ for negative $t(=-1)$.}
\end{table}

\begin{table}
\caption{The hole energy as a function of ${\bf k}$ and $J$ for positive
$t(=+1)$.}
\end{table}

\begin{table}
\caption{Total spin as a function of ${\bf k}$ and $J$ for positive $t(=+1)$.}
\end{table}

\newpage
\begin{center}
\begin{tabular}{||l|r|r||}   \hline
${\bf k}$&$\beta$  & $\alpha$\\ \hline
$(0,0)$             &0.112 & 1.056 \\ \hline
$(4\pi/3,0)$        &0.440 & 1.473 \\ \hline
$(\pi,\pi/\sqrt{3})$&0.517 & 0.610 \\ \hline
\end{tabular}
\end{center}
Table I

\newpage

\begin{center}
\begin{tabular}{||l|r|r|r||}   \hline
$J$ &     $A$ &    $B$ &    $C$ \\ \hline
0.05&-4.186 & 0.019 & 0.013 \\ \hline
0.1 &-4.018 & 0.032 & 0.025 \\ \hline
0.2 &-3.766 & 0.049 & 0.040 \\ \hline
0.3 &-3.585 & 0.070 & 0.058 \\ \hline
0.5 &-3.271 & 0.107 & 0.077 \\ \hline
\end{tabular}
\end{center}

Table II

\newpage

\begin{center}
\begin{tabular}{||l|r|r|r|r||}   \hline
$J$ ${\bf k}$& (0,0) &$(\pi,\pi/\sqrt{3})$&$(2\pi/3,0)$&$(4\pi/3,0)$\\ \hline
0.00 &-6.000 & -5.410 & -5.662 & -5.348 \\ \hline
0.05 &-5.258 & -4.856 & -5.040 & -4.807 \\ \hline
0.10 &-4.517 & -4.302 & -4.418 & -4.276 \\ \hline
0.20 &-3.368 & -3.342 & -3.349 & -3.382 \\ \hline
0.30 &-2.917 & -2.907 & -2.898 & -2.838 \\ \hline
0.50 &-2.368 & -2.195 & -2.454 & -2.065 \\ \hline
0.70 &-1.832 & -1.562 & -2.073 & -1.609 \\ \hline
0.80 &-1.567 & -1.359 & -1.895 & -1.483 \\ \hline
\end{tabular}
\end{center}

Table III
\newpage
\begin{center}
\begin{tabular}{||l|r|r|r|r||}   \hline
$J$ ${\bf k}$& (0,0) &$(\pi,\pi/\sqrt{3})$&$(2\pi/3,0)$&$(4\pi/3,0)$\\ \hline
0.00 &11/2 & 9/2 & 9/2 & 7/2 \\ \hline
0.05 &11/2 & 9/2 & 9/2 & 7/2 \\ \hline
0.10 &11/2 & 9/2 & 9/2 & 7/2 \\ \hline
0.20 & 7/2 & 5/2 & 5/2 & 5/2 \\ \hline
0.30 & 1/2 & 3/2 & 1/2 & 3/2 \\ \hline
0.50 & 1/2 & 3/2 & 1/2 & 3/2 \\ \hline
\end{tabular}
\end{center}

Table IV

\newpage
\begin{center}
\begin{tabular}{||l|r|r|r|r||}   \hline
$J$ ${\bf k}$& (0,0) &$(\pi,\pi/\sqrt{3})$&$(2\pi/3,0)$&$(4\pi/3,0)$\\ \hline
0.00 &-4.230 & -4.120 & -4.149 & -4.270 \\ \hline
0.05 &-4.076 & -3.928 & -3.971 & -4.192 \\ \hline
0.10 &-3.941 & -3.760 & -3.810 & -4.117 \\ \hline
0.50 &-2.964 & -2.706 & -2.699 & -3.569 \\ \hline
2.00 & 0.574 &  0.208 &  0.328 & -1.725 \\ \hline
15.0 &18.068 & 16.615 & 16.383 & 13.496 \\ \hline
\end{tabular}
\end{center}

Table V

\newpage

\begin{center}
\begin{tabular}{||l|r|r|r|r||}   \hline
$J$ ${\bf k}$& (0,0) &$(\pi,\pi/\sqrt{3})$&$(2\pi/3,0)$&$(4\pi/3,0)$\\ \hline
0.00 &3/2 & 1/2 & 3/2 & 1/2 \\ \hline
0.10 &3/2 & 1/2 & 1/2 & 1/2 \\ \hline
\end{tabular}
\end{center}

Table VI

\end{document}